\documentclass[11pt]{article} 
\usepackage{amsfonts,amscd,amsmath}

\def\NP#1#2{ Nucl.Phys. B#1 (#2)} 
\def\PL#1#2{ Phys.Lett. B#1 (#2)}
 
\def\MPL#1#2{ Mod.Phys.Lett.A#1 (#2)}

\def\HP#1#2{ JHEP #1 (#2)} 
\def\ap{ \alpha^{\prime}} 
\def\ai{\pi\alpha^{\prime}}
\def\pd{\partial}

\newcommand{\ep}{\text e}
\newcommand{\oh}{\frac{1}{2}}
\title{A Note on Open Strings in the Presence of Constant $B$-Field }
\author{Oleg Andreev\thanks{e-mail:  andreev@physik.hu-berlin.de}
\thanks{Permanent address: Landau Institute, Moscow, Russia}
\\
Humboldt--Universit\"at zu Berlin, Institut f\"ur Physik\\
Invalidenstra\ss e 110, D-10115 Berlin, Germany}
\begin{document} 
\date{}

\maketitle 
\begin{abstract} 
We consider the open string $\sigma$-model in the presence of a constant Neveu-Schwarz $B$-field 
on the world-sheet that is topologically equivalent to a disk with $n$ holes. First, we compute the 
$\sigma$-model partition function. Second, we make a consistency check of ideas about the appearance 
of noncommutative geometry within open strings.
\end{abstract}

\vspace{-10cm}
\begin{flushright}
hep-th/0001118       \\
HU Berlin-EP-00/08
\end{flushright}
\vspace{8.8cm}

The string $\sigma$-models 
 provide a basic tool to derive 
stringy low energy effective actions
 that contain all orders of the fields. 
One example  is the dilaton dependence 
of the effective action \cite{SS,WW,FTd}:
 the dilaton field is coupled 
to the Euler characteristic $\chi$ of the corresponding 
string world-sheet \cite{FTd} so that  each world-sheet 
contributes a factor $\ep^{-\chi\varphi}$ into the effective action. 
The second one is the 
Born-Infeld (BI) action which was derived  using  the open 
string $\sigma$-model on the 
disk \cite {FT,AC}  and adopting standard renormalization schemes.\footnote{For a 
 review  see \cite{T} and references therein.}
 Later it was realized that this action plays 
a crucial role in D-brane physics \cite{P,L,P1}. It is also worth mentioning that there are  known 
examples of string 
partition functions (one loop corrections to effective actions) for toroidal compactifications in the 
presence of constant metric and antisymmetric tensor which contain all orders of the fields \cite{Au}.

Recently, it was  pointed out  \cite{SW} 
that a   special renormalization scheme within
 the open string $\sigma$-model, 
a point splitting regularization, results
 in a rather peculiar situation where the space-time (brane) 
coordinates do not commute 
(see \cite{SW} and refs. therein). 
As a result, the low energy affective action 
becomes  a noncommutative BI action. Since 
different renormalization schemes should be  equivalent, there  
exists \cite{SW}  a change of variables ($\sigma$-model couplings) 
that relates the two BI actions.
The aim of this note is to  check  that the 
Seiberg-Witten relations are consistent at higher genus topologies. 

We study  the open string $\sigma$-model with a constant background metric and a constant 
Neveu-Schwarz 2-form  field on the world-sheet that is topologically equivalent to a  disk with $n$ holes. 
Such topologies appear in the perturbation theory of open orientable strings. 
Thus the world-sheet action is given by
\begin{equation}\label{ac}
S=\frac{1}{4\ai }\int_{\Sigma_n}d^2z\,\bigl(
g_{ij}\pd_aX^i\pd^aX^j-2i\ai B_{ij}\varepsilon^{ab}\pd_aX^i\pd_bX^j\bigr)+\chi\varphi
\quad,
\end{equation}
where $\Sigma_n$ means the disk with $n$ holes. $g_{ij},\,B_{ij},\,\varphi$ are the constant metric, 
antisymmetric tensor and dilaton fields, respectively. $X^i$ map the world-sheet to the brane and 
$i,j=1,\dots,p+1$. The world-sheet indices are denoted by $a,b$.

A natural object to compute is the $\sigma$-model partition function
\begin{equation}\label{pf}
Z_n[\varphi,g,B ]=\int [d q]_n\int {\cal D}X\,\ep^{-S}
\quad,
\end{equation}
where $[dq]_n$ means the modular measure for the disk with $n$ holes 
extended to arbitrary dimension. It proves irrelevant for what follows, so we do not explicitly write it 
down \footnote{See, e.g., \cite{MT} where it is written down for the annulus topology.}. 

To compute the partition function \eqref{pf} one can follow the approach  of \cite{FT} namely, first integrate 
over the internal points of the world-sheet to reduce the integral to the boundaries and then
split the 
integration variable $X^i$ on the constant $x^i$ and non-constant parts $\xi^i$. Since the action is 
quadratic in $\xi^i$ the problem is simply reduced to a computation of the corresponding functional 
determinant. The simplest case to consider is $n=0$ i.e., the path integral on the disk. In this case the problem 
is equivalent to the one solved in \cite{FT}. This is clear by   rewriting the term 
$B_{ij}\varepsilon^{ab}\pd_aX^i\pd_bX^j$ as a boundary interaction and replacing $B$ by 
$F$ \footnote{For the sake of simplicity, we use the matrix notations here and below.}. A subtle point 
we should mention here is due to a non-diagonal metric $g$. So, to get a GL(p+1) invariant answer, 
one must be careful with the measure of the integration
(see \cite{AMT}).
 Thus the partition function on the disk computed using  the $\zeta$-function 
regularization is given by
\begin{equation}\label{pf0}
Z_0[\varphi,g,B ] =
\ep^{-\varphi}\int [dx] \,\sqrt{\det g}\,\bigl[\det (1+2\ai g^{-1}B)\bigr]^{\oh}
\quad,
\end{equation}
where $[dx]=\frac{d^{p+1}x}{(2\ai )^{p+1}}$. The last factor is due to the integration 
over $\xi^i$. This  is clear within the perturbation theory where the $B$-term 
serves as an 
interaction.

Our aim now is to generalize the above result for arbitrary $n$. In fact,  what we actually need is only  
a generalization for the last factor in the integrand. Let us give simple, but a little bit heuristic, arguments 
that lead to a desired answer. It turns out that the problem has a simple solution 
in the framework of the so-called sewing operation for the world-sheets.  The latter is based on the idea of 
building surfaces by sewing together other ones. So let us begin with two disks and take a 
cylinder as a 
propagator between them. It is clear that the sewing operation produces a sphere. A crucial point 
here is that the partition function on the sphere does not depend on $B$. So, restricting ourselves to 
the $B$-dependence of the partition functions, namely $Z_{\text{sphere}}[B]\sim 1,\,
Z_0[B]\sim [\det (1+2\ai g^{-1}B)]^{\oh}$, etc., we have 
\begin{equation}\label{norm}
1\sim Z_{\text{sphere}}[B]\sim Z_0[B] \,\Pi [B] \,Z_0[B] 
\quad.
\end{equation}
As a result, we find the normalization of the propagator 
\begin{equation}\label{normP}
\Pi [B] \sim [\det (1+2\ai g^{-1}B)]^{-1}
\quad.
\end{equation} 
Now let us make a consistency check and compute the partition function on the annulus. This can be 
done at least in two ways. The first one is to sew its boundaries to get the torus topology. The second way is to 
sew it with two disks. The both ways lead to the same result 
\begin{equation}\label{check}
Z_1[B] \sim \det (1+2\ai g^{-1}B)
\quad.
\end{equation} 
Note that such a result was also found by direct
 calculation in \cite{CC}. 
  This is clear  by replacing 
$B\rightarrow F$ and rewriting the corresponding term in \eqref{ac} as boundary interactions. To be 
more precise, what we found corresponds to orientable non-planar diagrams for vector fields
(see also \cite{MT} where this case  corresponds to 
 the two field strengths at the two boundaries set to be opposite, $F_1=-F_2$).

It is now straightforward to get $Z_n[B] $. It is simply
\begin{equation}\label{zn}
Z_n[B] \sim [\det (1+2\ai g^{-1}B)]^{\frac{1+n}{2}}
\quad.
\end{equation} 
A crucial point here is that this factor does not depend on moduli.
Hence the 
partition function is given by 
\begin{equation}\label{pf-n}
Z_n[\varphi,g,B,\ap ]=\ep^{-\chi\varphi}\int [d q]_n\int [dx] \,\sqrt{\det g}\,
\bigl[\det (1+2\ai g^{-1}B)\bigr]^{1-\oh\chi}
\quad.
\end{equation} 
In above we have used the fact that the Euler characteristic $\chi$ 
of a planar disc  surface with $n$ holes is equal to $1-n$. 

Let us now give another way to derive the above result. The use of the point splitting regularization assumes 
that the metric $g$ becomes a new metric $G$ while all dependence on $B$ can be absorbed into 
the so-called star product that provides a multiplication law for other background fields.  In fact, in this case 
the action for the kinetic term is given by \cite{AD}
\begin{equation}\label{ac-n}
S=\frac{1}{4\ai }\int_{\Sigma_n}d^2z\,\ G_{ij}\pd_aX^i\pd^aX^j\ +\chi\hat\varphi
\quad,
\end{equation}
while interaction terms include the build in star products. Since there are no interaction terms
 in the problem at hand, the partition function should have a simple 
structure 
 due to standard dependence on  the dilatonic field and our definition of the measure. Thus we have 
\begin{equation}\label{pf-n-n}
\hat Z_n[\hat\varphi,G,\ap ]=\ep^{-\chi\hat\varphi}\int [d q]_n\int [dx] \,\sqrt{\det G}\,
\quad.
\end{equation} 
The new variables found by Seiberg and Witten in case of the disk topology are \cite{SW} 
\begin{equation}\label{nv}
\hat\varphi=\varphi+\oh\ln\det \bigl(G(g+2\ai B)^{-1}\bigr)
\quad ,\quad
G=(g-2\ai B)g^{-1}(g+2\ai B)
\quad.
\end{equation}
A simple algebra shows that the partition functions \eqref{pf-n} and \eqref{pf-n-n} coincide. So, the 
Seiberg-Witten relations \eqref{nv} hold on higher topologies too.

Finally, let us make some remarks. 

(i) First, let us remark that what we found can be reinterpreted in terms of vector fields. Indeed, we 
can consider a set of Abelian vector fields with constant field strengths $F^{(i)}$ such that different 
$A^{(i)}$'s coupled to different boundaries (in other words, take $n+1$ Wilson factors as interactions). 
A configuration of the $F$'s that allows to rewrite the boundary 
interactions as the bulk term exactly corresponds what we considered. From the physical point of view, 
such a configuration represents $n+1$ free Wilson factors with each factor contributing the Born-Infeld 
determinant.  

(ii) Second, it was conjectured in \cite{SW} that there exists a suitable regularization that interpolates 
between the Pauli-Villars ($\zeta$-function) regularization and the point splitting one. In the framework of 
the open string $\sigma$-model such a regularization was further developed in \cite{AD} where it was 
proposed to use the world sheet action 
\begin{equation}\label{ac-ps}
S=\frac{1}{4\ai }\int_{\Sigma_n}d^2z\,\bigl(
\tilde G_{ij}\pd_aX^i\pd^aX^j-2i\ai \Phi_{ij}\varepsilon^{ab}\pd_aX^i\pd_bX^j\bigr)+
\chi\tilde\varphi
\quad,
\end{equation}
where 
\begin{equation}\label{nv-ps}
\tilde\varphi =\varphi+\oh\ln\det\Bigl(\frac{\tilde G+2\ai\Phi}{g+2\ai B}\Bigr)
\quad,\quad
\tilde G=\Bigl(G^{-1}-\frac{1}{(2\ai )^2}\theta_0G\theta_0\Bigr)^{-1}
\quad,\quad
\Phi=-\frac{1}{(2\ai )^2}\tilde G\theta_0G
\quad.
\end{equation}
Here $\theta_0$ is a free matrix parameter. $G$ is given by Eq. \eqref{nv}.

Repeating the analysis that led us to Eq. \eqref{pf-n}, we can easily write down the partition function in 
the case of interest
\begin{equation}\label{pf-ps}
\tilde Z_n[\tilde\varphi,\tilde G,\Phi,\ap ]=\ep^{-\chi\tilde\varphi}\int [d q]_n\int [dx] \,
\sqrt{\det \tilde G}\,
\bigl[\det (1+2\ai \tilde G^{-1}\Phi)\bigr]^{1-\oh\chi}
\quad.
\end{equation} 
A simple algebra shows that $\tilde Z_n$ coincides with $Z_n$, so it also passes a consistency check on
higher genus topologies.

(iii)  It is not difficult to formally repeat the previous analysis  for superstring. To do so, it is more convenient 
to consider the NSR formalism within the point splitting regularization.
In other words, we add a set of the fermionic fields $\psi^i$ whose metric also is $G_{ij}$.  
It is simply to suggest what the superstring partition function should be. 
\begin{equation}\label{susy}
\hat\mathbf{Z}_n[\varphi, G,\ap ]=\ep^{-\chi\hat\varphi}\int [d \mathbf{q}]_n\int [dx] \,\sqrt{\det G}\,
\quad,
\end{equation}
where $[d \mathbf{q}]_n$ means a proper modular measure for superstring. Clearly, there is no problem  
with rewriting this expression in terms of $g,B$ and $\varphi$.

{\it Acknowledgements.} 
We would like to thank H. Dorn, R. Metsaev, and especially A.A. Tseytlin for useful discussions and 
critical comments.
This  work  is supported in part by the Alexander von Humboldt Foundation
and the European Community grant INTAS-OPEN-97-1312.



\end{document}